\documentclass[12pt, a4paper]{article}

\usepackage[inner=20mm,outer=20mm,top=25mm,bottom=25mm]{geometry}
\usepackage[onehalfspacing]{setspace}

\usepackage[english]{babel}
\usepackage{authblk}
\usepackage{amsmath, amssymb, physics}
\usepackage{cite, hyperref}

\newcommand{\be}{\begin{equation}}
\newcommand{\ee}{\end{equation}}
\newcommand{\bea}{\begin{eqnarray}}
\newcommand{\eea}{\end{eqnarray}}

\begin{document}	
	
	\thispagestyle{empty}
	\begin{center}
		
		\vspace*{2cm}
		{\LARGE \bf The $F$-term Problem and other}\\
		\vspace{0.7cm}
		{\LARGE \bf Challenges of Stringy Quintessence}\\
		\vspace{1.5cm}
		{\large Arthur Hebecker, Torben Skrzypek and Manuel Wittner}\\
		\vspace{0.5cm}
		\textit{Institut f\"ur Theoretische Physik, Ruprecht-Karls-Universit\"at Heidelberg, \\
			Philosophenweg 19, 69120 Heidelberg, Germany}\\
		\vspace{0.5cm}
		a.hebecker@thphys.uni-heidelberg.de, skrzypek@thphys.uni-heidelberg.de, wittner@thphys.uni-heidelberg.de
		\vspace{1cm}\\
		September 18, 2019
		
		\vspace{1cm}
		\textbf{Abstract}\\
	\end{center}
	
	\noindent 
	We attempt a systematic analysis of string-theoretic quintessence models as an alternative to metastable de Sitter vacua. It appears that, within the boundaries of what is known, large-volume type-IIB flux compactifications are
	preferred. Here the quintessence scalar is the ratio of certain 4-cycle volumes. It has already been noticed that the volume modulus, which must be stabilized, tends to remain too light. One may call this the ``light volume problem''. In addition, we identify an ``$F$-term problem'': The positive energy density of standard-model SUSY breaking is higher than the depth of all known negative contributions. We discuss what it would take to resolve these issues and comment on partially related challenges for axionic quintessence. In particular, large cancellations between positive and negative potential terms appear unavoidable in general. As a further challenge, one should then explain why a small de-tuning cannot be used to uplift into a deep slow-roll regime, violating de Sitter swampland conjectures.	
	\newpage
	
	\section{Introduction}
	Stabilizing all moduli of a 4D string compactification, especially in the presence of supersymmetry (SUSY) breaking and positive cosmological constant, is notoriously difficult. Already the simplest realistic models~\cite{Kachru:2003aw,Balasubramanian:2005zx} involve several ingredients and significant tuning. As a result, some skepticism concerning these models may be justified (see~\cite{Bena:2009xk, McOrist:2012yc, Dasgupta:2014pma, Bena:2014jaa, Quigley:2015jia, Cohen-Maldonado:2015ssa, Junghans:2016abx, Moritz:2017xto, Sethi:2017phn, Danielsson:2018ztv, Moritz:2018sui, Cicoli:2018kdo, Kachru:2018aqn, Kallosh:2018nrk, Bena:2018fqc, Kallosh:2018psh, Hebecker:2018vxz, Gautason:2018gln, Heckman:2018mxl, Junghans:2018gdb, Armas:2018rsy, Gautason:2019jwq, Blumenhagen:2019qcg, Bena:2019mte, Dasgupta:2019gcd} for a selection of papers criticizing and defending de Sitter constructions). Recently, this has culminated in the proposal of a no-go theorem against stringy quasi-de Sitter constructions. Concretely, in the single-modulus case, this includes the claim that~\cite{Obied:2018sgi,Garg:2018reu,Ooguri:2018wrx}
	\be
	\abs{V'}\geq c\cdot V \qquad \text{or}\qquad V''\leq - c'V\,, \label{conjecture}
	\ee
	where $c$ and $c'$ are order-one numbers.\footnote{
	We set $M_{\rm P}=1$ except in equations with units and when its explicit appearance enhances readability.} This may be taken as an incentive to better understand the KKLT and Large-Volume-Scenario (LVS) constructions and improve on them (see \cite{Hamada:2018qef, Kallosh:2019oxv, Hamada:2019ack, Carta:2019rhx, Kachru:2019dvo} for progress in refuting some of the criticism based on 10D considerations). However, it is also interesting to take the opposite perspective: Accept the above de Sitter swampland conjecture as true and see what would be left of string phenomenology.
	
	The most direct way out has already been emphasized in~\cite{Obied:2018sgi, Agrawal:2018own}: The presently observed cosmic acceleration would have to come from a stringy version of quintessence \cite{Wetterich:1987fm,Peebles:1987ek,Caldwell:1997ii}.\footnote{
		For the purpose of this paper we are generous concerning the parameter $c$, allowing it to be significantly smaller than unity to match experimental restrictions~\cite{Agrawal:2018own, Heisenberg:2018yae, Akrami:2018ylq, Raveri:2018ddi}.
	}
	The latter is, however, not easy to realize (see e.g.~\cite{Hellerman:2001yi, Chiang:2018jdg, Cicoli:2018kdo, Marsh:2018kub, Han:2018yrk, Acharya:2018deu, Hertzberg:2018suv, vandeBruck:2019vzd, Baldes:2019tkl} for discussions). The most promising candidates for stringy quintessence are moduli (see e.g.~\cite{Cicoli:2012tz, Olguin-Tejo:2018pfq, Emelin:2018igk}) and axions (see e.g.~\cite{Nomura:2000yk, Svrcek:2006hf, Panda:2010uq, Chiang:2018jdg, Cicoli:2018kdo, DAmico:2018mnx, Ibe:2018ffn}), which  are both ubiquitous in string compactifications. 
	In the present paper, we attempt to make progress not so much towards providing an explicit model but at least towards carefully specifying the challenges that have to be overcome. Our focus will be on ultra-light K\"ahler moduli in type IIB flux compactification, following the most explicit examples available~\cite{Cicoli:2011yy,Cicoli:2012tz}. We will postpone comments on axion quintessence to section 5.
	
	Quintessence models rely on a scalar slowly rolling down a potential. Cosmology constrains its mass, which we define as $\sqrt{V''}$, to be smaller than the Hubble scale: $\abs{m_\phi}\lesssim H_0\approx10^{-33}\text{ eV}\sim\order{10^{-60}}M_{\rm{P}}$ \cite{Tsujikawa:2013fta}. This lightness makes the quintessence scalar susceptible to fifth-force constraints, ruling out in particular the overall-volume modulus. Our main candidates will hence be ratios of certain 4-cycle volumes.
	
	Stringy quintessence needs large hierarchies between the mass of the quintessence scalar, the volume-modulus mass, and the mass scale of Standard-Model (SM) superpartners. In the spirit of~\cite{Cicoli:2011yy, Cicoli:2012tz}, we use a large volume $\mathcal{V}$ and an anisotropic geometry to suppress the loop corrections which make the quintessence scalar massive. However, this also lowers the mass scale of the volume modulus, leading to what we want to call the ``light volume problem''.
	
	Moreover, even if some new effect making the volume sufficiently heavy could be established (see~\cite{Cicoli:2011yy, Cicoli:2012tz} for suggestions), another problem remains: The SM-superpartner masses induced by the available K\"ahler modulus $F$-terms are too low. This can be overcome by introducing a dedicated SUSY-breaking sector on the SM brane. Yet, even taking the corresponding mediation and hence $F$-term energy scale as low as possible, a significant uplifting effect on the full scalar potential is induced. We call this the ``$F$-term problem''. In the given setting, the corresponding energy density is comparable to the positive and negative energy scales cancelling each other in the underlying no-scale model and much above the residual $1/{\cal V}^3$ AdS-potential of the LVS stabilization mechanism.
	
	The rest of the paper is structured as follows: We introduce the phenomenological requirements in section 2 and translate them to model-building restrictions in section 3, where we re-derive the light volume problem. In section 4 we present the $F$-term problem arising from the phenomenologically required SUSY breaking. A discussion of possible loopholes, axion quintessence and alternative approaches follows in section 5 before we conclude in section 6.
	
	\section{Preliminaries and Requirements}
	We will focus on compactifications of type IIB string theory on Calabi-Yau orientifolds with O3/O7 planes. One reason is that this setting is particularly well-studied and has proven to be phenomenologically promising (see~e.g.~\cite{Kachru:2003aw, Balasubramanian:2005zx, Giddings:2001yu, Denef:2008wq}). A closely related reason is the no-scale structure arising after the flux stabilization of complex-structure moduli. This allows one to go to a large volume and make use of different small corrections to the K\"ahler-moduli scalar potential. As we will see, this appears to be precisely what one needs for the large hierarchies required in the present context.
	
	The 4D effective theory arising at the classical level is characterized by ${\cal N}=1$ supergravity (SUGRA) with K\"ahler and superpotential
	\be
	K_{\rm{tot}}=-2\ln{\cal V}(T+\bar{T})+K_{\rm{cs}}(z,\bar{z})\qquad \mbox{and} \qquad W=W(z)\,.
	\ee
	Here $T$ stands symbolically for all K\"ahler moduli and $z$ for the complex-structure moduli together with the axio-dilaton. After solving the $F$-term equations $D_zW=(\partial_z+K_z)W=0$, by which the $z$-moduli get stabilized, one ends up with
	\be
	K=-2\ln{\cal V}(T+\bar{T})\qquad \mbox{and} \qquad W=W_0=\,\mbox{const.}\,,
	\ee
	where we have absorbed any additive constants in $K$ into a redefinition of $W$. Since the volume ${\cal V}$ is a homogeneous function of degree 3/2 of the K\"ahler moduli $T=\{T_1,T_2,\cdots\}$, the scalar potential vanishes identically,
	\be
	V=e^K(K^{i\bar{\jmath}}D_i WD_{\bar{\jmath}}\bar{W}-3\abs{W}^2)=K_{i\bar{\jmath}}F^i \bar{F}^{\bar{\jmath}}-3e^K\abs{W}^2=0\,.\label{noscale}
	\ee
	This no-scale structure breaks down due to quantum corrections, giving
	\be
	V=\delta V_{\rm{np}}+\delta V_{\alpha'}+\delta V_{\rm{loop}}\neq 0\,,\label{corrs}
	\ee
	where one distinguishes:
	\begin{itemize}
		\item \textbf{Non-perturbative corrections} due to D7-brane gaugino condensation or E3-brane instantons. While they generically correct both K\"ahler and superpotential, their main effect on the scalar potential comes from $W\,\,\to \,\, W=W_0+A_ie^{-a_iT^i}\,.$
		\item \textbf{$\alpha'$ corrections}, which arise from higher-order terms in the 10D action. The established leading effect~\cite{Becker:2002nn} can be accounted for by $K\,\,\to\,\, K=-2\ln(\mathcal{V}+\xi)\,.$
		\item \textbf{String-loop corrections}, which can also be viewed as field-theoretic loop corrections in a Kaluza-Klein (KK) compactification and would naively affect the K\"ahler potential more strongly than the $\alpha'$ corrections: $K\,\,\to\,\, K+\delta K_{\rm{loop}}\,.$ However, due to an extended no-scale cancellation, their effect on the scalar potential is subdominant~\cite{vonGersdorff:2005bf, Berg:2005ja, Berg:2005yu,Cicoli:2007xp}.
	\end{itemize}

	At large volume, the terms in (\ref{noscale}) scale as $1/{\cal V}^2$ and the no-scale structure may be viewed as an exact cancellation of scalar potential terms at this order. The terms in (\ref{corrs}) are suppressed by further volume powers, as we will discuss in more detail below. As a result, K\"ahler moduli are parametrically light at large ${\cal V}$, which makes them natural candidates for the quintessence scalar. Conversely, the extreme lightness of quintessence enforces ${\cal V}\gg 1$.
	
	Possibilities for including the SM are fractional D3-branes at a singularity or D7-branes wrapping a 4-cycle~\cite{Conlon:2005ki}. In the best-understood examples, this will give rise to a SUSY version of the SM. SUSY will then have to be broken at least at about 1~TeV$\, \sim 10^{-15}M_{\rm{P}}$. 
	
	With this general setting fixed, we proceed by listing the phenomenological requirements, to be justified momentarily:
	\begin{enumerate}
		\item \textbf{Light quintessence modulus} $\phi$ with $m_{\phi}\lesssim 10^{-60}M_{\rm{P}}\,.$
		\item \textbf{Heavy superpartners} with $m_S\gtrsim 10^{-15} M_{\rm{P}}\,.$
		\item \textbf{Heavy KK scale} with $m_{\rm{KK}}\gtrsim 10^{-30}M_{\rm{P}}\,.$
		\item \textbf{Heavy volume modulus} with $m_{\mathcal{V}}\gtrsim 10^{-30} M_{\rm{P}}\,.$
	\end{enumerate}
	The first two requirements are obvious from what has been said above: the need for a slowly rolling scalar and consistency with the LHC. The third requirement follows from the fact that standard 4D Newtonian gravity has been tested at scales below $0.2$~meV~$\sim$~$1$~mm$^{-1}$\cite{Kapner:2006si}.
	
	Finally, the fourth requirement is obtained if one notices that, after compactification, the Ricci scalar of the 4D theory obtains a prefactor ${\cal V}$. Then, after Weyl rescaling to the 4D Einstein frame, the scalar field corresponding to ${\cal V}$ couples to matter fields (both from D3 and D7 branes) with approximately gravitational strength. However, such fifth-force effects are ruled out by the very same experiments that test gravity at the sub-millimeter scale \cite{Damour:2010rp, Acharya:2018deu, Kapner:2006si} (measuring the Eddington parameter in the post-Newtonian expansion). Hence the volume modulus must be sufficiently heavy.
	
	Comparing the first and last requirement, it is immediately clear that $\phi$ cannot be the volume modulus. It can, however, be one of the K\"ahler moduli measuring the relative size of different 4-cycles. We will see below that, while these can be much lighter than ${\cal V}$, reaching the extreme level of $10^{-60}M_{\rm{P}}$ proves non-trivial. We also note that such K\"ahler moduli couple to matter, though not as strongly as ${\cal V}$. These couplings tend to violate the equivalence principle, forcing them to remain about a factor of $10^{-11}$ below gravitational strength~\cite{Damour:2010rp}. Fifth-force constraints on stringy quintessence models have recently been studied in detail in~\cite{Acharya:2018deu}, where a lower bound on the compactification volume, which suppresses the couplings to other K\"ahler moduli, was found for a number of models. Our focus in this paper is different and concerns the more elementary issue of mass hierarchies in the scalar potential and the SUSY-breaking scale. The volume needed for these hierarchies is in general even larger than prescribed by the bounds from fifth-force constraints.
	
	\section{Mass Hierarchies and resulting Bounds}
	
	As explained, we focus on K\"ahler moduli and rely on the corrections of (\ref{corrs}) to generate a non-zero potential. It will hence be useful to recall their generic volume-scaling (e.g. from \cite{Cicoli:2009zh}). In doing so, we suppress all ${\cal O}(1)$ coefficients and write $\tau^i:=\frac{1}{2}(T^i+\bar{T}^i)$:
	\be
	\delta V_{\rm{np}}\sim \frac{\sqrt{\tau_\text{s}}{\rm{e}}^{-2a_s\tau_\text{s}}}{\cal V}+\frac{W_0 \tau_\text{s} {\rm{e}}^{-a_s\tau_\text{s}}}{{\cal V}^2}\,\,\to\,\,\frac{W_0^2}{{\cal V}^3}\log^{3/2}(W_0/{\cal V}) \,,\qquad
	\delta V_{\alpha'}\sim\frac{W_0^2}{{\cal V}^3}\,,
	\qquad \delta V_{\rm{loop}}\sim \frac{W_0^2}{{\cal V}^{10/3}}\,.
	\label{ecorr}
	\ee
	Naively, the non-perturbative correction is always subleading due to its exponential suppression. However, it may be relevant if it is induced by a `small cycle' $\tau_\text{s}$. In this case, after the modulus $\tau_\text{s}$ is integrated out, a volume-dependent effect arises which (up to a log-enhancement) scales in the same way as the $\alpha'$ correction. The interplay of these two effects may then provide the celebrated volume stabilization in LVS \cite{Balasubramanian:2005zx,  Conlon:2005ki, Cicoli:2009zh} with an AdS minimum at ${\cal V}={\cal V}_0$ and 
	\be
	V_{\rm{LVS}}\,\,\sim\,\, \delta V_{\rm{np}}+\delta V_{\alpha'}\,\,\sim \,\,\frac{W_0^2}{{\cal V}_0^3}\,.\label{LVS}
	\ee
	Here ${\cal V}_0$ can be exponentially large, with the exponent being $\sim \chi^{2/3}/g_\text{s}$ (where $\chi$ is the Euler characteristic of the Calabi-Yau and $g_\text{s}$ the string coupling).
	
	As explained before, this is exactly what we need: The volume must be very large but stabilized at a sufficiently high scale to avoid fifth-force constraints. Crucially, even though ${\cal V}={\cal V}(T)$ is in general a complicated function of all K\"ahler moduli, $V_{\rm{LVS}}$ depends only on the overall volume. The role of quintessence can then be played by any combination of K\"ahler moduli other than the overall volume (and excluding any `small cycles' -- i.e.~those for which $\exp(-\tau)$ is not negligibly small). 
	
	We now need to discuss moduli masses in more detail. First, $\tau_\text{s}$ (and similar moduli stabilized by their non-perturbative corrections) are heavy: $m_{\tau_\text{s}}\sim W_0/{\cal V}$. We will not discuss them any further and also neglect their contributions to the volume. In the moduli space of the remaining `large cycles' $T^i$, one direction (corresponding to the overall volume ${\cal V}$) is stabilized by the non-perturbative and $\alpha'$ corrections. The other moduli receive a mass from $V_{\rm loop}$. Although also other corrections could contribute to the moduli masses, as for example the poly-instanton corrections in \cite{Cicoli:2011yy}, we will only discuss loop corrections here, since they generally contribute to any modulus and thus provide a lower limit on moduli masses. To discuss them, we focus on the submanifold defined by ${\cal V}=\,$const.~and, in addition, ignore the axions. The kinetic term is then defined by the metric $K_{i\bar{\jmath}}=K_{ij}$, restricted to that submanifold. After canonical normalization of the kinetic terms the moduli masses are obtained from the second-derivative matrix of the scalar potential $\partial_i \partial_{\bar{\jmath}}V$. The specific structure of $K_{ij}$ for large-cycle volumes allows one to estimate the masses simply by the square root of the relevant potential term (see the appendix and \cite{Skrzypek} for more details). This also holds for the volume modulus so that,
	according to \eqref{ecorr} (see also \cite{Conlon:2005ki}), one finds parametrically
	\be
	m_{\mathcal{V}}\sim\sqrt{\delta V_{\alpha'}}\sim\frac{W_0}{\mathcal{V}^{3/2}}\,\,,\qquad m_{\mathcal{\tau }^i}\sim m_\phi \sim\sqrt{\delta V_{\rm{loop}}}\sim\frac{W_0}{\mathcal{V}^{5/3}}\,.\label{masses}
	\ee
	Here we use the notation $m_\phi$ since we already know that the quintessence field $\phi$ will be one of those large-cycle volumes (more precisely volume ratios) present in addition to ${\cal V}$. 
	
	Combining (\ref{masses}) with the required scales listed in the previous section, one finds
	\be
	\order{10^{30}}\lesssim\frac{m_{\mathcal{V}}}{m_{\phi}}\sim\mathcal{V}^{1/6}\qquad \Rightarrow \qquad\mathcal{V}\gtrsim\order{10^{180}}\,.
	\label{iso1}
	\ee
	This is a very large volume and will result in very small KK scales given by
	\be
	m_{\rm{KK}}=\frac{M_{\rm{s}}}{R}\sim\frac{M_{\rm{P}}}{\mathcal{V}^{1/2+1/6}}\lesssim\order{10^{-120}}M_{\rm{P}}\,,
	\label{iso2}
	\ee
	which is in conflict with requirement 3. Here we have used that the string scale $M_{\rm{s}}$ of the 10D Einstein frame is given by $M_{\rm{s}}=M_{\rm{P}}/\sqrt{\mathcal{V}}$ and the typical Radius $R$ of the compactification is the sixth root of the volume, assuming isotropy.
	
	The loop corrections involving the quintessence modulus thus have to be suppressed more strongly than by $\mathcal{V}^{-10/3}$. As suggested in~ \cite{Cicoli:2011yy, Cicoli:2012tz}, anisotropic compactifications may provide the required suppression. To understand this idea, a heuristic argument for the power of $-10/3$ in the loop corrections is useful~\cite{Cicoli:2009zh, Cicoli:2011yy}: From a 4D point of view, loop corrections arise from loops of all light fields below a cutoff $\Lambda$, where the 4D description breaks down. This $\Lambda$ is assumed to be given by the lowest KK scale, where the theory becomes effectively higher-dimensional.\footnote{
	This is a non-trivial assumption since loop corrections may, of course, also arise in higher-dimensional field theory or directly at the string level. In fact, one probably has to assume that the restoration of a sufficiently high level of SUSY above the KK scale cuts off the loop integrals. However, in the present case SUSY is broken by fluxes, and these penetrate not just the large-radius but {\it all} extra dimensions. So further scrutiny may in fact be required to justify the use of the {\it lowest} KK scale as a cutoff.}
	The fields running in the loops contribute with different masses and signs and the potential at 1-loop order will be the SUSY analogue of the Coleman-Weinberg potential \cite{Coleman:1973jx, Ferrara:1994kg}: 
	\be
	V=V_\text{tree}+\frac{1}{64\pi^2}\text{STr}\mathcal{M}^0\cdot\Lambda^4\log\frac{\Lambda^2}{\mu^2}+\frac{1}{32\pi^2}\text{STr}\mathcal{M}^2\cdot\Lambda^2+\frac{1}{64\pi^2}\text{STr}\mathcal{M}^4\log\frac{\mathcal{M}^2}{\Lambda^2}+...\,.
	\label{cw}
	\ee
	The second term disappears due to SUSY. The third term involves the supertrace $\text{STr}\mathcal{M}^2$ of all fields running in the loops. In general 4D $\mathcal{N}=1$ SUGRA, this supertrace is given by $\text{STr}\mathcal{M}^2=2Qm_{3/2}^2$, where $Q$ is a model dependent $\order{1}$ coefficient, while $m_{3/2}$ is the gravitino mass given by $\abs{W}/\mathcal{V}$. This allows us to estimate the lowest order loop corrections by
	\be 
	\delta V_{\text{loop}}\sim Am_{\rm{KK}}^2m_{3/2}^2+Bm_{3/2}^4\sim A m_{\rm{KK}}^2 \frac{W_0^2}{\mathcal{V}^2}+ B\frac{W_0^4}{\mathcal{V}^4}
	\label{potential}
	\ee
	with $\order{1}$ constants $A$ and $B$.\footnote{Although the terms in \eqref{potential} could in principle cancel each other, we will not discuss cancellations here and refer to the discussion.} As discussed earlier, in an isotropic compactification the first term gives exactly the familiar $\mathcal{V}^{-10/3}$ dependence which results in too small KK scales. Therefore, we now assume an anisotropic compactification with $l$ large dimensions of radius $R\sim\mathcal{V}^{1/l}$ and the other $6-l$ dimensions at string scale for highest possible suppression. This creates a hierarchy between the KK scales so that the heavy KK modes have masses at string scale while the light ones have masses of order $m_{\rm{KK}}\sim\mathcal{V}^{-(1/2+1/l)}$. Looking only at the first term in \eqref{potential}, we observe that smaller $l$ makes the quintessence field lighter. However, this improvement ends when the value of the first term falls below that of the second, $m_{\rm{KK}}$-independent term. This occurs at $l=2$, which is hence the optimal value on which we now focus. We note that further suppression can apparently be achieved if $l=1$ and, in addition, $W_0$ is tuned small. But, as we will explain below, this does not resolve the problems we will face.
	
	Thus, in the anisotropic scenario with $l=2$, the quintessence scalar gets loop corrections only at order $\mathcal{V}^{-4}$ which in contrast to \eqref{masses} induces a quintessence mass\footnote{
	We again refer to the appendix for a justification of the formula $m_\phi\sim\sqrt{\delta V_{\rm loop}}$.}
	\be
	m_\phi \sim\sqrt{\delta V_{\rm{loop}}}\sim\frac{W_0}{\mathcal{V}^2}\,.\label{suppressed}
	\ee
	Since requirement 3 bounds the volume to $\mathcal{V}\lesssim \order{10^{30}}$ we can marginally source the right quintessence mass. However, using $m_{\mathcal{V}}$ from \eqref{masses} and $m_\phi$ from \eqref{suppressed} together with our phenomenological requirements 1 and 4, we conclude
	\be
	\order{10^{-30}}\gtrsim\frac{m_\phi}{m_{\mathcal{V}}}\sim\mathcal{V}^{-1/2}\sim m_{\rm{KK}}^{1/2}\qquad \Rightarrow \qquad \order{10^{-60}}\gtrsim m_{\rm{KK}}\label{h1}\,,
	\ee 
	where in the last step, we see a contradiction with requirement 3 arising as the KK scale becomes too low.
	So even in the anisotropic case the required hierarchy cannot be achieved through the standard LVS approach.\footnote{As mentioned above, we can further suppress $V_\text{loop}$ by choosing $l<2$ and tuning $W_0$ small. The obvious possibility is $l=1$ corresponding to one large and five small dimensions. One may also consider more complicated geometries where several radii between $1/M_{\rm{s}}$ and some maximal radius $1/M_{\rm{KK}}$ are used. This latter case may be treated by using an effective $l$ with $1 \leq l \leq 2$ in the crucial formula for $m_{\rm{KK}}$. Either way, repeating the analysis which led to \eqref{h1} one arrives at $m_{\rm{KK}} \leq \mathcal O (10^{-30-15l})$ for general $l$. Thus, requirement 3 is always violated and the light volume problem cannot be resolved by going to $l\leq2$.}
	
	We will refer to this problem, which has already been noted in~\cite{Cicoli:2011yy, Cicoli:2012tz}, as the ``light volume problem''. To resolve it, one needs an extra contribution to the scalar potential, which gives the volume modulus a higher mass. This is already critical. However, as we will see momentarily, things get even more challenging if we take into account SUSY breaking. This will provide an independent argument for a new scalar-potential term, fixing also its sign and prescribing a significant overall magnitude.
	
	\section{The $F$-term Problem}
	
	It is necessary to ensure that the SM superpartners are sufficiently heavy (requirement 2). This will prove to be very challenging. For instance, the gaugino mass is given by
	\be
	m_{1/2} = \frac{1}{2}\frac{F^m \partial_m f}{\text{Re} f},
	\ee	
	 where $f$ is the gauge-kinetic function. If the SM gauge group is realized on D7-branes, $m_{1/2}$ scales as $|W|/\mathcal{V}$. For D3 realizations, the soft scale is suppressed more strongly~\cite{Conlon:2005ki} -- so this does not help. Due to the aforementioned phenomenological requirements 1 and 2, the hierarchy between the quintessence field and the gaugino must fulfill
	\be
	\frac{m_\phi}{m_{1/2}} \lesssim \mathcal{O}(10^{-45}).
	\ee
	We can furthermore use the first term in (\ref{potential}) to conclude that $m_\phi\gtrsim m_{\rm{KK}}m_{3/2}$ and observe that $m_{3/2}\sim m_{1/2}$ in the present setting. This implies $m_\phi/m_{1/2} \gtrsim m_{\rm{KK}}$, in conflict with requirement 3. We conclude that the gaugino mass cannot be generated by the SUSY breaking of the K\"ahler moduli alone.
	
	Instead, to obtain large enough gaugino masses, we need a further source of SUSY breaking. One can realize this on the SM brane through mediation from a hidden sector where SUSY is broken spontaneously by the non-vanishing $F$-term of a spurion field $X$. Without loss of generality, we will use the language of spontaneous SUSY breaking even in the case that this breaking is realized locally (at the same Calabi-Yau singularity) and directly at the string scale.\footnote{
		In this case one may speak of non-linearly realized SUSY (see~\cite{Ferrara:2016een} for recent progress in this context). One may, however, also continue to use the language of e.g.~$F$-term SUSY breaking in SUGRA, sending the masses of the fields in the SUSY-breaking sector to infinity.
	}
	
	According to \cite{Conlon:2005ki}, the moduli $X_{\alpha}$ of D3-branes enter the K\"ahler potential $K(T+\overline{T})$ through the replacement
	\be
	2\tau^i=T^i+\bar{T}^{\bar{\imath}}\quad\to\quad2\tau'^i=T^i+\bar{T}^{\bar{\imath}}+k^i(X^\alpha,\bar{X}^{\bar{\alpha}})\,,
	\ee
	where $k^i(X^\alpha,\bar{X}^{\bar{\alpha}})$ are some real-valued functions. These may be chosen quadratic or higher-order since any linear components can be absorbed into the definition of the $T^i$ or removed via a K\"ahler transformation. We will call the resulting new K\"ahler potential $K'$. Now computing the scalar potential involves inverting a $2\times 2$ block matrix, with the blocks corresponding to the $T^i$ or $X^\alpha$ variables. One finds that the $F$-term contribution from the K\"ahler moduli cancels against the gravitational term $-3{\rm{e}}^{K'}\abs{W}^2$ in standard no-scale fashion, leaving behind a term\footnote{
		Here 
		we assume that $X=0$ in the vacuum. To be completely explicit, one may think of $k\sim X\overline{X}-a(X\overline{X})^2$ and $W=bX$ in the single-field case.
	}
	\be
	V\supset \delta V_X=K'_{\alpha\bar{\beta}}F_X^\alpha \bar{F}_X^{\bar{\beta}} \qquad{\rm{where}}\quad K'_{\alpha\bar{\beta}}=K_i\partial_\alpha\partial_{\bar{\beta}}k^i\,,\quad F_X^\alpha={\rm{e}}^{K'/2}K'^{\alpha\bar{\beta}}\partial_{\bar{\beta}}\bar{W}\,.
	\ee	
	
	Thus, SM-brane SUSY breaking gives a positive contribution to the scalar potential, which is added on top of the zero potential resulting from the K\"ahler-moduli no-scale structure. Now consider a simple toy model with a single spurion field $X$ and $F$-term $F_X\equiv F$. Let SUSY breaking be mediated  through higher-dimension operators suppressed by $M$, which we define to be the mediation scale of the flat SUSY limit (see \cite{Skrzypek} for details). After canonical normalization of $X$ and its $F$-term, one has $m_{1/2}\sim F/M$
	(and similarly for the other soft terms), which implies
	\be
	\delta V_X\sim F^2\sim M^2m_{1/2}^2\,.
	\label{fterm}
	\ee
	
	In the D7-brane case, a similar substitution, $S+\bar{S}\to S+\bar{S}+k(X,\bar{X})\,,$ is applied to the dilaton term in $K$. Since the dilaton $S$ is stabilized by fluxes it can be treated as a constant, so the scalar potential is simply $\abs{D_XW}^2$. This generates the positive $F$-term even more directly so we will not discuss this case separately.
	
	Soft masses are phenomenologically constrained to be at least $\sim\,$TeV$\,\sim\order{10^{-15}}M_{\rm{P}}$. Moreover, $M$ should be high enough to hide the SUSY-breaking sector. It is then natural to assume $M\gtrsim\order{10^{-15}}M_{\rm{P}}\,,$\footnote{We will more carefully exclude lower values in Section \ref{sec:limits_on_delta_V}.}
	which implies $\delta V_X \sim M^2m_{1/2}^2 \sim \order{10^{-60}}M_{\rm{P}}^4\,.$ This is of the same order of magnitude as the cancellation in the standard no-scale scenario, i.e.~far larger than the first-order LVS corrections.\footnote{
		Indeed, 
		as noted earlier $m_\phi\gtrsim m_{\rm{KK}}m_{3/2}$ so that the canceling terms in the no-scale potential are of order $V_{\rm no-scale}\sim m_{3/2}^2\lesssim m_\phi^2/m_{\rm{KK}}^2\lesssim 10^{-60}M_{\rm P}^4\,,$ where we enforce requirements 1 and 3.}
	Thus $\delta V_X$ raises the height of the scalar potential to very large positive values which cannot be canceled by the terms in $V_{\rm{LVS}}$ of \eqref{LVS}.
	
	\subsection{Limits on $\delta V_X$} \label{sec:limits_on_delta_V}
	
	Since $\delta V_X$ has emerged as a key issue for the most popular stringy quintessence models, we want to evaluate more carefully whether this hidden-sector contribution to the scalar potential can be consistently tuned to smaller values. Recall from \eqref{fterm} that it scales as $\delta V_X \sim m_{1/2}^2 M^2$. Since the gaugino mass should not be smaller than $\order{10^{-15}} M_{\rm{P}}$, the only option is to reduce $M$ and $F$ at the same time.
	
	While explicit model building is not our main goal, we note in passing that realistic scenarios with small $F$-terms and correspondingly small mediation scale are not easy to get. For successful constructions in the 5D context and a discussion of the problems one encounters see~\cite{Dimopoulos:2014aua, Dimopoulos:2014psa, Garcia:2015sfa}.
	
	A simultaneous reduction of $M$ and $F$ implies a reduction of the gravitino mass. In the past, there have been many investigations that aimed at constraining the latter using data from electroweak colliders \cite{Brignole:1997sk,Luty:1998np, Abbiendi:2000hh,Heister:2002ut,Achard:2003tx,Abdallah:2003np, Antoniadis:2012zz,Mawatari:2014cja} like LEP or hadronic ones \cite{Brignole:1998me,Acosta:2002eq,Klasen:2006kb,deAquino:2012ru,Aad:2015zva} like the Tevatron. These bounds on $m_{3/2}$ translate into lower limits of the SUSY-breaking scale, which typically constrain $\sqrt{F}$ to be larger than a few $100 \, \text{GeV}$.
	
	The most recent and stringent bounds result from missing-momentum signatures in $pp$ collisions at the LHC. To understand the emergence of such bounds, let us consider an exemplary toy model where SUSY is spontaneously broken in a hidden sector through a non-vanishing $F$-term in the vacuum and mediated to the SM sector via the interaction terms
	\be
	\mathcal{L}_\text{int} = \frac{a}{M^2} \int d^4 \theta X^\dagger X \Phi^\dagger \Phi + \frac{b}{M} \int d^2 \theta X W^\alpha W_\alpha + \text{h.c.}\,, \label{eq:lagrangian_interaction}
	\ee
	where $\Phi$ is a chiral superfield representing quarks $q$ and squarks $\tilde{q}$ whereas $W^\alpha$ is the supersymmetric field-strength tensor of a vector superfield $V$ representing gluons $g$ and gluinos $\tilde{g}$. A non-zero $F$ in the vacuum will generate soft masses for the squarks and gluinos, which are given by $m_{\tilde{q}}^2 = a F^2/M^2$ and $m_{\tilde{g}} \sim b F/M$, respectively. The hidden-sector field $X$ contains the goldstino $\tilde{G}$, which gets eaten by the gravitino due to the super-Higgs mechanism. In the limit $\sqrt{s}/m_{3/2} \gg 1$, the helicity-1/2 modes dominate over the helicity-3/2 modes and, according to the gravitino-goldstino equivalence theorem \cite{Casalbuoni:1988kv, Casalbuoni:1988qd}, yield the same S-matrix elements as the goldstinos. Hence in this simple discussion, we identify the gravitino with the goldstino. We are now interested in processes which turn two hadrons into a hadronic shower plus gravitinos, where the latter induce a missing-momentum signature. For instance, we can consider the process of two quarks in the initial state and two gravitinos in the final state with a gluon being eradiated from one of the initial quarks, resulting in a hadronic shower. The gluon radiation costs a factor $\sqrt{\alpha_S}$. Several beyond-SM processes contribute to the crucial  $qq$-$\tilde{G}\tilde{G}$-amplitude. One of them is the direct 4-particle coupling from \eqref{eq:lagrangian_interaction}:
	\be
	\sim \frac{a}{M^2} \bar{\tilde{G}} \tilde{G} \bar{q} q \subset \frac{a}{M^2} \int d^4 \theta X^\dagger X \Phi^\dagger \Phi\,.
	\ee
	Due to the prefactor $a/M^2$, this vertex contributes a factor $1/F^2$ to the amplitude so that the cross section will be proportional to $\alpha_S/F^4$. This $F^{-4}$-dependence of the cross section is typical 
	for such processes and therefore the upper limits on them, provided by measurements at hadron colliders, translate into lower bounds on $F$.
	
	In a recent experimental analysis of the ATLAS collaboration \cite{Aad:2015zva}, the process $pp \rightarrow \tilde{G} + \tilde{q}/\tilde{g}$ is considered, whereupon the squark or gluino decays into a gravitino and a quark or gluon, respectively. Depending on the squark and gluino masses, as well as on their ratios, the authors derive lower bounds on the gravitino mass around $m_{3/2} \approx (1 - 5) \times 10^{-4} \, \text{eV}$ corresponding to SUSY-breaking scales $\sqrt{F} \approx (650 - 1460) \, \text{GeV}$.
	
	In \cite{Maltoni:2015twa}, not only the process $pp \rightarrow \tilde{G} + \tilde{q}/\tilde{g} \rightarrow 2 \tilde{G} + q/g$ but also direct gravitino-pair production with a quark or gluon emitted from the initial proton as well as squark or gluino pair production with a following decay into gravitinos and quarks or gluons are considered. Taking into account all three processes, the authors of \cite{Maltoni:2015twa} use the model-independent 95\% confidence-level upper limits by ATLAS \cite{ATLAS:2012zim} on the cross section for gravitino + squark/gluino production to constrain $\sqrt{F} > 850 \, \text{GeV}$. This is done for the case when the squark and gluino masses are much larger than those of the SM particles so that they can effectively be integrated out (in the paper, the value $m_{\tilde{q}/\tilde{g}} = 20 \, \text{TeV}$ is used). In other scenarios, where one or both of these two types of superpartners have lower masses, the bound becomes even higher.
	
	We conclude that, in accordance with the current experimental status, the mass scale of SUSY breaking $\sqrt{F}$ cannot be lowered significantly below $100 \, \text{GeV} - 1 \, \text{TeV}$ so that $\delta V_X$ can be at most a few orders of magnitude below $\order{10^{-60}}M_{\rm{P}}^4$. Such a contribution cannot be canceled by any known term in our scenario as has been discussed already.
	
	\subsection{Need for a new contribution}
	
	We have seen that requirement 2 of heavy superpartners implies the presence of a large positive contribution $\delta V_X$ to the scalar potential. This would raise the potential far above the observed energy density $\order{10^{-120}}M_{\rm{P}}^4$, rendering this whole scenario unviable. Since we do not know how to avoid this effect, it appears logical to assume the presence of a further negative contribution of equal magnitude, which fine-tunes $V$ to a level consistent with observations. In the preferred case of $l=2$ and for $W_0\sim{\cal O}(1)$, the required magnitude is $\delta V_{\text{new}}\sim\mathcal{V}^{-2}$. Such a contribution may also solve the light volume problem \eqref{h1}. Indeed, if its volume dependence is generic, one expects an induced volume-modulus mass $m_{\mathcal{V}}\sim\mathcal{V}^{-1}$. This is just enough to build all required hierarchies. 
	
	We emphasize that this contribution is substantially hypothetical and that the nature of its generation and form is not understood. Possible effects suggested in \cite{Cicoli:2011yy,Cicoli:2012tz} are loop corrections from open strings on the SM brane and the back-reaction of the bulk to the brane tension along the lines of the SLED models \cite{Burgess:2004xk}. Open string loops may induce a Coleman-Weinberg potential with cutoff at the string scale $M_{\rm{s}}\sim M_{\rm{P}}/\sqrt{\mathcal{V}}$, such that the leading term scales as $M_{\rm{s}}^4\sim M_{\rm{P}}^4/\mathcal{V}^2$. Although this is the correct order of magnitude for $\delta V_{\text{new}}$, the volume dependence appears to be too simple to allow for volume-modulus stabilization. Moreover, being a higher-order correction to the brane sector, we would assume it to already be part of the low-energy effective K\"ahler potential for $X$ and the SM fields which we used to derive $F$-terms and induce superpartner masses. As such it could not contribute the required negative energy to cancel the critical $F$-term.
	
	As mentioned above, a counteracting contribution could also be found in the bulk back-reaction. Since the SM-brane tension is the origin of the large $F$-term, a back-reaction to this tension from the bulk appears to be promising. Still, as our analysis shows, it remains a challenge to include this in the 4D effective theory, specifically in the 4D effective SUGRA, which we expect to arise at low energies in the string theoretic settings we consider (see also \cite{Nilles:2003km,Burgess:2005wu, Burgess:2011mt,Cicoli:2011yy, Cicoli:2012tz} for related discussions).
	
	Finally, in the context of the de Sitter swampland conjecture \eqref{conjecture}, our $F$-term implies yet another difficulty. Even if the new term $\delta V_{\text{new}}$ cancels the $F$-term to leave a sufficiently small potential, a small change in the SM or SUSY-breaking parameters can raise the $F$-term and with it the residual scalar potential to violate the conjecture. This is also problematic in other models and we will come back to this issue in the following sections. 
	
	\section{Loopholes and alternative Approaches}
	
	There are several potential loopholes in our analysis. The first one is the possibility that the quintessence modulus is extremely light (i.e.~the loop-induced potential is extremely flat) by fine-tuning.\footnote{For example, one could imagine a model where the two terms in \eqref{potential} cancel to a very small residue.}
	However, this seems implausible for the following reason:
	The flatness must hold on a time scale of order $H_0^{-1}$. In quintessence models which respect the de Sitter conjecture \eqref{conjecture}, the scalar field has to run sufficiently far during such a period. Indeed, from the Klein-Gordon-equation in Friedmann-Robertson-Walker background together with $|V'|/V\lesssim 1$ it follows that $\Delta\phi\sim\order{1}$ in one Hubble time.
	In a Taylor expansion of $\delta V_{\rm{loop}}$, we therefore have to take into account all orders of $\Delta\phi$. It is thus not enough to fine-tune $\delta V_\text{loop}$ at one point but we must tune an infinite number of derivatives to small values. This cannot be coincidental but has to be based on some mechanism or symmetry. Although in our specific model such a perfect decoupling of one K\"ahler modulus from the loop corrections seems implausible, there might of course be other constructions where the required sequestering can be achieved (see \cite{Acharya:2018deu, Heckman:2019bzm} for discussions).

	Another possibly critical point is the approximation of loop corrections through the Coleman-Weinberg potential \eqref{cw} with $m_{\rm{KK}}$ as a cutoff. Here, one has to be concerned that no other, stronger corrections arise. This seems possible, for example, since the KK scale is far below the weak scale. Thus, when applying the formula, one has to do so in a setting where the SM~brane (with SUSY broken at a higher scale) has already been integrated out. This needs further scrutiny. Another concern is that even in the bulk SUSY may not be fully restored above $m_{\rm{KK}}$ due to the effect of bulk fluxes. 
	Still, we trust the formula to at least give a lower bound on loop corrections that cannot be neglected and thus makes our conclusions inevitable.
	
	A number of alternative approaches to quintessence building from string theory have been proposed.
	Let us first comment on the possibility of axion quintessence. Based on the SUGRA scalar potential, one generically expects an axion potential
	\be
	V=\Lambda^4\cos\left(\frac{\phi}{f}\right)+a\,, \qquad \Lambda^4\sim M_{\rm{P}}^2m_{3/2}^2\rm{e}^{-S_{\rm inst.}}\,.
	\label{axion}
	\ee 
	This could provide the required dark energy if $\phi$ is at the ``hilltop'' and, at the same time, satisfy the second condition of \eqref{conjecture} (assuming reasonably small $c'$). For simplicity, let us start the discussion taking $a=0$. Then the slow-roll condition, which we need phenomenologically, requires a trans-Planckian axion decay constant $f$~\cite{Panda:2010uq}. But this is in conflict with quantum-gravity expectations or, more concretely, the weak gravity conjecture for axions~\cite{Banks:2003sx, ArkaniHamed:2006dz}: 
	\be
	f\leq\order{1}M_{\rm{P}}\qquad\mbox{or}\qquad S_{\rm{inst.}}\leq \alpha \frac{M_{\rm{P}}}{f}\,.
	\ee 
	The conflict is strengthened if one recalls that the potential must be tiny, i.e. $M_{\rm{P}}^2m_{3/2}^2{\rm{e}}^{-\alpha M_{\rm{P}}/f}\lesssim 10^{-120}M_{\rm{P}}^4$. For $\alpha \sim\order{1}$, this implies $f\sim\order{10^{-2}}M_{\rm{P}}$, which is in conflict with slow-roll. As suggested in \cite{Cicoli:2018kdo}, one might hope to ease the tension by employing the constant contribution $a$ to the potential \eqref{axion}.\footnote{Another idea to resolve the conflict would be to move away from the hilltop to a point in field space where both slow-roll conditions are as weak as possible. This turns out not to work.} If $a$ is negative, the slow-roll condition is violated even more strongly. Positive $a$ greater than $\Lambda^4$ leads to a violation of the de Sitter conjecture at the minimum. The best option is then $a=\Lambda^4$ which, however, does not help much: The slow-roll requirements on $f$ change only by a factor $\sqrt{2}$, so $f$ still needs to be at the Planck scale.
	
	With this naive approach we would have to violate the weak gravity conjecture by assuming an unacceptably large $S_{\rm inst.}$. However, the weak gravity conjecture is presumably on stronger footing than the de Sitter conjecture, so this is against the spirit of the swampland discussion. Instead, alternative elements of model building may be invoked to save axion quintessence. An option is the use of axion monodromy~\cite{Panda:2010uq}. Another idea developed and discussed in \cite{Nomura:2000yk, delaFuente:2014aca, Hebecker:2017uix, Ibe:2018ffn, Hebecker:2019vyf} is a further suppression of the prefactor of the axion potential. A specific model with a highly suppressed axion potential for an electroweak axion has been developed in \cite{Nomura:2000yk, Ibe:2018ffn}. We note that the most obvious suppression effects are related to high-quality global symmetries in the fermion sector, suggesting a relation between the weak gravity conjecture and global-symmetry censorship~\cite{Hebecker:2019vyf,Fichet:2019ugl}.
	
	If such models succeed in providing a sufficiently flat potential, we still have to account for large enough SUSY breaking in the full model to generate heavy SM superpartners. The large $F$-term required has to be canceled to allow for the flat axion potential to dominate. Assuming this cancellation to be implemented, we can again slightly change the SUSY-breaking contributions to shift the axion potential to positive values and violate the de Sitter conjecture at the minima. The full model would need to balance out these changes by some intricate mechanism.
	
	An alternative approach to building a quintessence potential from KKLT-like ingredients has been taken in \cite{Emelin:2018igk} where the quintessence field is given by the real part of a complexified K\"ahler modulus. This K\"ahler modulus runs down a valley of local axionic minima in the real direction. Since the universe is assumed to be in a non-supersymmetric non-equilibrium state today, it can evolve at positive potential energies. However, since the potential has to be sufficiently small to constitute a quintessence model, the superpotential has to be tuned to very small values, which results in a small gravitino mass. It appears that one needs further SUSY breaking and the $F$-term problem re-emerges. 
	
	An interesting alternative to quintessence has been introduced in \cite{Hardy:2019apu}: The zero-temperature scalar potential is assumed to satisfy the de Sitter conjecture, but a thermally excited hidden sector stabilizes a scalar field at a positive-energy hilltop. The authors illustrate this idea using a simple Higgs-like potential $V=-m_\phi^2\phi^2/2+\lambda\phi^4+C$. Since the hidden sector must not introduce too much dark radiation, the temperature and hence also $m_\phi$ are bounded from above by today's CMB temperature, which is roughly $0.24$~meV. 
	Since this model does not need an approximate no-scale structure to ensure an extremely flat potential at large ${\cal V}$, our $F$-term problem does not immediately arise. 
	
	However, it makes an indirect appearance as follows: Both the present toy model potential as well as more general models of this type are expected to have a minimum somewhere. In the present case, its depth is $m_\phi^4/16\lambda$, which is very small unless $\lambda$ is truly tiny. Now, since some $F$-term effect $\delta V_X$ must be present somewhere in the complete model, a small de-tuning of this $\delta V_X$ will be sufficient to lift the model into the swampland. Thus, some form of conspiracy must again be at work for this model to describe our world and the de Sitter conjecture to hold simultaneously. 
	
	A way out is provided by assuming that $\lambda\sim \order{10^{-64}}$ and available $\delta V_X$ are bounded at $\sim\,$TeV. Then the minimum is too deep to be lifted to de Sitter by de-tuning. Even then, one has to be careful to ensure that $|V''|/V$ does not become too small as one uplifts the model by de-tuning the SUSY-breaking effect. We approximate the possible de-tuning by the order of magnitude of the $F$-term itself: $\Delta(\delta V_X)\sim\delta V_X\sim F^2$. As a result $|V''|/\Delta(\delta V_X) \sim m_\phi^2/F^2\sim \order{(10^{-31})^2/10^{-60}} \sim \order{10^{-2}}$, which is critical in view of the de Sitter conjecture. Thus, even in this rather extreme case, a version of the $F$-term problem can at best be avoided only marginally.
	
	\section{Conclusion}
	
	We have analyzed stringy quintessence on the basis of the phenomenologically required hierarchies between quintessence mass, volume-modulus mass, SUSY-breaking scale and KK scale. Within the type IIB framework, one is naturally led to the setting of~\cite{Cicoli:2012tz}, where quintessence corresponds to the rolling in K\"ahler moduli space at fixed overall volume. One also immediately notices the light volume problem, which requires a new ingredient (see~\cite{Cicoli:2011yy} for a suggestion) to make the volume modulus sufficiently heavy.
	
	In addition, we have identified what one might call an $F$-term problem. It derives from the fact that SUSY-breaking by the $F$-terms of K\"ahler moduli is far to weak phenomenologically. Thus, an additional SUSY-breaking sector on the SM brane is required. This generates a sizable uplift contribution to the scalar potential. The well-known negative contributions associated with $\alpha'$-, loop and non-perturbative effects are much too small to cancel this uplift, given that we are at very large values of the volume modulus. 
	
	The situation can then be summarized as follows: The construction of quintessence from a K\"ahler modulus in Type IIB flux compactifications requires a yet unknown contribution to the scalar potential. This is not only needed to stabilize the volume modulus but, in addition, it must be negative and of the order $\delta V_{\text{new}}\sim\mathcal{V}^{-2}$ to compensate for the effect of SUSY breaking. Moreover, this correction may not raise the mass of the other K\"ahler moduli.
	
	Finally, if the above requirements can be met, a further issue arises: In the framework envisioned above, today's tiny vacuum energy is the result of a precise cancellation between the SM-related $F$-term uplift and $\delta V_{\text{new}}$. It would then appear that models with a slightly higher $F$-term 
	uplift, induced by a tiny change in the SM or SUSY-breaking sector parameters, should also exist. Such models would have an unchanged tiny slope $V'$ but a much higher potential $V$, violating even a mild form of the de Sitter swampland conjecture (such as (\ref{conjecture}) with a fairly small $c$ and $c'$). 
	
	Possibilities to go forward include the specification and study of the missing potential effect $\delta V_{\text{new}}$, the construction of models which completely evade the effective-4D-SUGRA logic that we used, or the study of entirely different string-theoretic settings. The latter may, for example, use type IIA or the heterotic framework or appeal to different quintessence candidates, like the rolling towards large complex structure or small string coupling. Of course, in the first case one may find oneself at large volume after all, as suggested by mirror symmetry. In the second case, one faces the risk that the string scale falls below the KK scale. Returning to our analysis in this paper, we suspect that in many cases some variant of our $F$-term problem, rooted in the strong SUSY breaking in the SM, is likely to be relevant.

	\section*{Acknowledgements}
	
	We would like to thank Pablo Soler and Michele Cicoli for fruitful discussions. This work is supported by the Deutsche Forschungsgemeinschaft (DFG) under Germany's Excellence Strategy EXC-2181/1 - 390900948 (the Heidelberg STRUCTURES Excellence Cluster). Furthermore, M.W. thanks the DFG for support through the Research Training Group ``Particle Physics beyond the Standard Model'' (GRK 1940).

		\section*{Appendix: Estimating Moduli Masses from the Potential}
	
	We will argue that under reasonable assumptions the mass scale of a physical modulus is usually set by the highest order term $\delta V$ in the scalar potential that involves the respective modulus: 
	\be
	m^2\gtrsim \delta V\,.
	\ee	
	This is easy to see for the volume modulus but requires justification for the other moduli. Although heavier masses can easily arise for `small-cycle' moduli which correspond to small terms in ${\cal V}$, much lighter masses require some kind of cancellation, which will generally involve tuning. 
		
		To illustrate the idea, consider the toy model lagrangian
	\be
	\mathcal{L}=\frac{\partial_\mu X \partial^\mu X}{2X^2} + V(X)\,,\qquad\mathrm{where}\qquad V''(X)\sim\frac{V(X)}{X^2}\,.
	\ee  	
	The canonical field is introduced through $X=\exp(\phi)$. Then the physical mass squared is the second derivative of the potential w.r.t. $\phi$. Given our assumption about $V''(X)$, this is of the same order of magnitude as the potential itself. Thus, suppressing $\order{1}$ coefficients, the approximation $m^2\sim \delta V$ is justified. 

	For the volume modulus the argument is basically as in the toy model above. So we now restrict our attention to the submanifold of constant ${\cal V}$ in the space of real moduli $\tau^1,...,\tau^n$. We choose an arbitrary trajectory on this submanifold and parameterize it as
	\be
	(\tau^1(\phi),...,\tau^n(\phi))=(\tau^1(0)\mathrm{e}^{\xi^1(\phi)\phi},...,\tau^n(0)\mathrm{e}^{\xi^n(\phi)\phi})\,.
	\ee
	We normalize our parameter $\phi$ so that it takes the value $0$ at the point of interest $\tau^i\equiv \tau^i(0)$. The coefficient vector $\xi^i\equiv \xi^i(0)$ is chosen to be $\order{1}$ valued.
	Now the lagrangian for motion along the trajectory contains the kinetic term 
	\be
	\mathcal{L}\supset \mathcal{L}_{\mathrm{kin}}=\sum_{ij}K_{ij}\tau^i\tau^j\xi^i\xi^j \partial_\mu \phi \partial^\mu \phi\,.
	\ee
	We can compute the K\"ahler metric from the K\"ahler potential $K=-2\ln(\mathcal{V}(\tau^i))$ and since we are moving along the submanifold of constant volume we can use
	\be
	\sum_{i}\mathcal{V}_i\tau^i\xi^i=0 \qquad \text{such that} \qquad \mathcal{L}_{\mathrm{kin}}=-2\sum_{ij}\frac{\mathcal{V}_{ij}}{\mathcal{V}}\tau^i\tau^j\xi^i\xi^j \partial_\mu \phi \partial^\mu \phi\,.
	\ee
	Unless there is significant cancellation between terms in $\mathcal{V}$ we can assume 
	\be
	\mathcal{V}_{ij}\lesssim \frac{\mathcal{V}}{\tau^i\tau^j}
	\label{heur}
	\ee
	and since $\xi^i$ was chosen $\order{1}$, the whole prefactor of $\partial_\mu \phi \partial^\mu \phi$ can be assumed to be $\order{1}$ or smaller. A small prefactor can arise from a small contribution in $\mathcal{V}(\tau^i)$ as for example in the standard LVS example of $\mathcal{V}=\tau_\text{b}^{3/2}-\tau_\text{s}^{3/2}$ where $\tau_\text{s}$ is a small modulus and gets a small prefactor in the kinetic term.  The canonical normalization will thus either not change or even increase the order of magnitude of the modulus mass. 

	Turning to the potential, we see that, since we move along the submanifold, any contribution only involving the volume does not contribute to the mass, as for example $V_{\rm{LVS}}$ in \eqref{LVS}. Turning to the leading-order contribution $\delta V$ involving the other moduli (in our case string-loop corrections) we will rewrite the potential in the coordinates $(\mathcal{V}, \tau^1,...\tau^{n-1})$ where we have solved the constraint of staying on the submanifold for a suitable $\tau^n$. We introduce indices $k$ and $l$ which run over $\{1,...,n-1\}$ in contrast to $i$ and $j$. The mass squared of our modulus is now determined by the Hessian of the potential contracted with the vector $\delta\tau^k$ corresponding to an infinitesimal shift in $\phi$ : 
	\begin{equation}
	m^2\sim \delta V_{kl}\frac{\delta\tau^k}{\delta\phi}\frac{\delta\tau^l}{\delta\phi}=\sum_{kl}\delta V_{kl}\tau^k\tau^l\xi^k\xi^l\sim\order{\delta V}\,.
	\end{equation}
	Here we have to assume that after rewriting the potential in terms of $(\mathcal{V}, \tau^1,...\tau^{n-1})$ it is still sufficiently well behaved to allow for an order of magnitude estimate $\delta V_{kl}\sim\delta V/\tau^k\tau^l$, resembling \eqref{heur}. Since the choice of trajectory was arbitrary, we assume a similar scaling for all moduli involved except for the volume modulus. Bearing in mind the possible mass enhancement from the canonical normalization, we estimate
	\be
	m^2\gtrsim \delta V\,.
	\ee
	We note that the requirements are met in many simple cases, for example the models of \cite{Cicoli:2011yy, Cicoli:2012tz}. A more detailed analysis can be found in \cite{Skrzypek}.	
		
	\bibliography{References}
	\bibliographystyle{JHEP}
\end{document}